\documentclass[aps,prd,preprint,showpacs,showkeys]{revtex4}

\usepackage{latexsym}
\usepackage{graphicx}
\usepackage{subfigure}
\begin{document}

%\preprint{arXiv:yymm.nnnn [hep-th]}

\title{Lifshitz scalar, brick wall method, and GUP in Ho\v{r}ava-Lifshitz Gravity}

%%%%%%%%%%%%%%%%%%%% PRD Author Style
\author{Myungseok Eune}
\email[]{younms@sejong.ac.kr}
\affiliation{Institute of Fundamental Physics, Sejong University, Seoul 143-747, Korea}

\author{Wontae Kim}
\email[]{wtkim@sogang.ac.kr}
\affiliation{
  Center for Quantum Spacetime, Sogang University, Seoul 121-742, Korea
  \\
  Department of Physics, Sogang University, Seoul, 121-742, Korea \\
  Korea Institute for Advanced Study, Seoul, 130-722, Korea}

\date{\today}

\begin{abstract}
  Using the brick wall method, we study statistical entropy for
  spherically symmetric black holes in Ho\v{r}ava-Lifshitz gravity. In
  particular, a Lifshitz scalar field is considered in order to
  incorporate foliation preserving diffeomorphism which eventually
  gives a modified dispersion relation. Finally, we obtain the area
  law without UV cutoff for $z > 3$, and discuss some of consequences
  in connection with the generalized uncertainty principle.
\end{abstract}



\maketitle

\section{Introduction}
\label{sec:intro}

% HL gravity
Recently, Ho\v{r}ava has put forward a renormalizable theory of
gravity when the scaling dimensions of space and time are different,
which is called Ho\v{r}ava-Lifshitz (HL) gravity \cite{Horava:2008jf,
  Horava:2009uw}.  It is power-counting renormalizable for $z = d$ and
superrenormalizable for $z>d$, where $z$ and $d$ are scaling dimension
and number of spatial dimensions, respectively.  Subsequently, there
have been extensive studies for the HL gravity \cite{Blas:2009qj,
  Wang:2009azb, Lu:2009em, Cai:2009pe, Cai:2009ar, Colgain:2009fe,
  Kehagias:2009is, Myung:2009va, Cai:2010ud, Koutsoumbas:2010pt,
  Kiritsis:2009sh, Brandenberger:2009yt, Kiritsis:2009vz,
  Majhi:2009xh, Son:2010qh}, such that various black
holes~\cite{Lu:2009em, Cai:2009pe, Cai:2009ar, Colgain:2009fe,
  Kehagias:2009is, Myung:2009va, Cai:2010ud, Koutsoumbas:2010pt} and
cosmological solutions \cite{Kiritsis:2009sh, Brandenberger:2009yt,
  Kiritsis:2009vz, Majhi:2009xh, Son:2010qh} have been intensively
studied. Moreover, it has been claimed that the nonisotropic scaling
of spacetime related to the foliation preserving diffeomorphism (FPD)
gives a modified dispersion relation \cite{Romero:2009qs,
  Rama:2009xc}.  As for the nonisotropic scaling, even in a flat
spacetime, it generically leads to an intriguing dispersion relation
of the form, $E^2 - c^2 \big[p^2 + \cdots + \Lambda_z (p^2)^z \big] =
m^2 c^4$, where $c$, $m$, and $\Lambda_z$ are the speed of light, the
mass of a particle, and a parameter, respectively~\cite{Rama:2009xc}.
Of course, it can be generalized in non-flat spacetimes. There have
been some studies for modified dispersion relations in black hole
physics, similarly to this dispersion relation~\cite{Corley:1996ar,
  Jacobson:2007jx}.

% Entropy - brick wall
On the other hand, it has been known that the entropy of a black hole
is proportional to the area of its event horizon. For calculating the
statistical entropy, the brick-wall method suggested by {'}t Hooft can
be used~\cite{thooft}, where the cutoff parameter should be introduced
to handle the divergence near the event horizon. Since degrees of
freedom of a field are dominant near the horizon, the brick wall can
be replaced by a thin layer or a thin spherical box~\cite{hzk}.  By
the way, the cutoff parameter located just outside the horizon can be
avoided if we consider the generalized uncertainty principle (GUP)
\cite{li:plb02540, kkp, zrl, Ko:2010zz}. Actually, the mode counting
can be done from the horizon to the minimal length and it gives finite
density because of the modification of phase space volume and the
dispersion relation.

% our work
In this paper, we would like to study the statistical entropy of
spherically symmetric black holes in the HL gravity using the
(thin-layered) brick wall method. For this purpose, we introduce a
Lifshitz scalar field rather than the usual scalar field to
incorporate the nonisotropic symmetry up to the matter sector. In this
semi-classical calculation, the resulting entropy shows that for $z >3
$ corresponding to the super renormalizable case of HL gravity, the
ultraviolet (UV) cutoff parameter can be avoided, so that a thin layer
can be located just outside the horizon similarly to the case of
GUP\@. Assuming the Bekenstein-Hawking entropy which is proportional
to the area of horizon, we can naturally fix the size of thin layer
depending only on the scaling.  In section~\ref{sec:bh}, we
recapitulate the Ho\v{r}ava-Lifshitz gravity and black hole
solutions. In section~\ref{sec:dispersion}, WKB approximations with
the modified dispersion relation for the Lifshitz scalar field will be
considered. In section~\ref{sec:entropy:HL}, the statistical entropy
will be given by counting the number of quantum states, and find the
condition to give the area law of entropy.  In
section~\ref{sec:z:even}, some issues related to the modified
dispersion relation will be presented. Finally, some discussions will
be given in section~\ref{sec:discussion}.

\section{Black holes in HL gravity}
\label{sec:bh}

We briefly review HL gravity in a self-contained manner, and introduce
black hole solutions for (3+1)-dimensional HL gravity. On general
grounds, as like the ADM decomposition of the metric in Einstein
gravity, the (3+1)-dimensional metric can be decomposed into
\begin{equation}
  \label{metric:HL}
  ds^2 = - N^2 c^2 dt^2 + g_{ij} (dx^i + N^i dt) (dx^j + N^j dt),
  \qquad i,j = 1,2,3
\end{equation}
where $N$ and $N^i$ are the usual lapse and shift functions. An
anisotropic scaling transformation of time $t$ and space $\vec{x}$ is
given by
\begin{equation}
  \label{transf:scale}
  t \to b^z t, \qquad x^i \to b x^i,
\end{equation}
under which $g_{ij}$ and $N$ are invariant, while $N^i \to b^{1-z}
N^i$. We have scaling dimensions given by $[t] = z$, $[x^i] = 1$, and
$[c]=[N^i] = 1-z$ in units of spatial length.

The kinetic action in the Ho\v{r}ava-Lifshitz gravity is given
by~\cite{Horava:2009uw}
\begin{equation}
  \label{action:K}
  I_K = \frac{2}{\kappa^2} \int dt d^3x \sqrt{g} N \left[ K_{ij}
    K^{ij} - \lambda K^2 \right],
\end{equation}
where $\kappa^2$ and $\lambda$ are a coupling related to the Newton
constant $G_N$, and an additional dimensionless coupling constant,
respectively. The extrinsic curvature is given by $K_{ij} =
\frac{1}{2N} (\dot{g}_{ij} - \nabla_i N_j - \nabla_j N_i)$, where the
overdot denotes the derivative with respect to time $t$ and $\nabla_i$
is the covariant derivative with respect to spatial metric
$g_{ij}$. Note that the original kinetic part of Einstein-Hilbert
action can be recovered when $\lambda=1$ and $\kappa^2 = 32 \pi
G_N/c^2$.  Moreover, the power-counting renormalizability requires
$z\ge3$. Now, the potential term of action is determined by the
``detailed balance condition'' as follows~\cite{Horava:2009uw}:
\begin{equation}
  \label{action:V}
  I_V = \frac{\kappa^2}{8} \int dt d^3 x \sqrt{g} N E^{ij}
  \mathcal{G}_{ijkl} E^{kl},
\end{equation}
where $E^{ij}$ comes from 3-dimensional relativistic action in the
form of
\begin{equation}
  \label{tensor:E}
  E^{ij} = \frac{1}{\sqrt{g}} \frac{\delta W[g_{ij}]}{\delta g_{ij}},
\end{equation}
and the generalized De Witt metric $\mathcal{G}^{ijkl}$ and its
inverse metric $\mathcal{G}_{ijkl}$ for $\lambda \ne 1/3$ are given by
\begin{eqnarray}
  & &\mathcal{G}^{ijkl} = \frac12 (g^{ik} g^{jl} + g^{il} g^{jk}) -
  \lambda g^{ij} g^{kl},   \label{DWmetric} \\
  & &\mathcal{G}_{ijkl} = \frac12 (g_{ik} g_{jl} + g_{il} g_{jk}) -
  \frac{\lambda}{3\lambda - 1} g^{ij} g^{kl}, \quad \mathrm{for}\ \lambda
  \ne \frac13, \label{inverseDW}
\end{eqnarray}
with the normalization condition of
\begin{equation}
  \label{def:inverseDW}
  \mathcal{G}^{ijkl} \mathcal{G}_{klmn} = \frac12 (\delta^i_m
  \delta^j_n + \delta^i_n \delta^j_m).
\end{equation}
In particular, the relativistic acton $W$ is expressed as $W = W_1 +
W_2$ for $z=3$ \cite{Horava:2009uw, Kiritsis:2009sh}, where $W_1$ and
$W_2$ are given by
\begin{eqnarray}
  W_1 &=& \mu \int d^3x \sqrt{g} (R - 2\Lambda_W), \label{W1} \\
  W_2 &=& \frac{1}{w^2} \int d^3x \sqrt{g} \varepsilon^{ijk}
  \Gamma^m_{il} \left(\partial_j \Gamma^l_{km} + \frac23 \Gamma^l_{jn}
    \Gamma^n_{km} \right) \label{W2}
\end{eqnarray}
with $\varepsilon_{ijk} = \sqrt{g} \epsilon_{ijk}$ and $\epsilon_{123}
= 1$.  Here, $\mu$ and $w^2$ are coupling constants and $\Lambda_W$ is
a cosmological constant.

Let us assume that the line element of a spherically symmetric black
hole can be written in the form of
\begin{equation}
  \label{metric:symmetric}
  ds^2 = -f \tilde{N}^2  c^2 dt^2 + \frac{dr^2}{f} + r^2 (d\theta^2 +
  \sin^2\theta\, d\phi^2),
\end{equation}
where $f=f(r)$ and $\tilde{N} = \tilde{N}(r)$. For an arbitrary
$\lambda$ for $z=3$, there are three solutions \cite{Lu:2009em}.  The
first one is given by
\begin{equation}
  \label{bh:LMP:1st}
  f = 1 - \Lambda_W r^2,
\end{equation}
with an arbitrary function $\tilde{N}$. The others for $\lambda>1/3$
are given by
\begin{eqnarray}
 f &=& 1 - \Lambda_W r^2 - \alpha (\sqrt{-\Lambda_W} r)^{2\lambda \pm
    \sqrt{6\lambda - 2}/(\lambda-1)}, \label{bh:LMP:2nd:f}\\
  \tilde{N} &=& (\sqrt{-\Lambda_W} r)^{-(1+3\lambda \pm 2 \sqrt{6\lambda
    - 2})/(\lambda - 1)},   \label{bh:LMP:2nd:N}
\end{eqnarray}
where $\alpha$ is an integration constant. On the other hand, $\lambda
= 1$ for $z=3$, the asymptotically flat solution along with a
vanishing cosmological constant is given by \cite{Kehagias:2009is}
\begin{eqnarray}
  f &=& 1 + \omega r^2 - \sqrt{r(\omega^2 r^3 + 4\omega
    M)}, \label{bh:KS:f} \\
  \tilde{N} &=& 1, \label{bh:KS:N}
\end{eqnarray}
where $\omega = 16\mu^2/\kappa^2$ and $M$ is an integration
constant. Moreover, in the modified Ho\v{r}ava-Lifshitz gravity
proposed in Ref.~\cite{Blas:2009qj}, the other types of spherically
symmetric solutions have been also studied~\cite{Kiritsis:2009vz}.  In
fact, we need not consider specific forms of black hole solutions as
long as the spherical symmetric ansatz holds since we shall calculate
the statistical entropy near the horizon without loss of generality.

\section{Modified and reduced dispersion relations}
\label{sec:dispersion}

We consider a complex scalar field $\varphi$ obeying the modified
Klein-Gordon equation implemented by FPD, which is assumed to be
\begin{equation}
  \label{eom}
  - \frac{1}{Nc\sqrt{g}} \partial_t \left(\frac{\sqrt{g}}{Nc}
    D_t \varphi \right) +
  \frac{1}{Nc} \nabla_i \left(\frac{N^i}{Nc} D_t\varphi \right)  - \left[
  \Lambda_0 + \Lambda_1 (-\nabla^2) + \cdots + \Lambda_z (-\nabla^2)^z
\right] \varphi = 0,
\end{equation}
where the derivative $D_t$ is defined by $D_t = \partial_t - N^i
\partial_i$, the Laplacian is given by $\nabla^2 \equiv g^{ij}
\nabla_i \nabla_j$, and the constants $\Lambda_n$ will be fixed
later. Note that Eq.~(\ref{eom}) may be induced from a certain action
especially for a constant lapse function, there appears such a
consideration in Ref.~\cite{Suyama:2009vy}. Now, applying WKB
approximation to Eq.~(\ref{eom}) with $\varphi = \exp[iS(t,x^i)]$, we
obtain a modified dispersion relation,
\begin{equation}
  \label{WKB}
  \frac{1}{N^2c^2} (p_t - N^i p_i)^2 - \left[ \Lambda_0 +
    \Lambda_1 p^2 + \cdots + \Lambda_z (p^2)^z \right]= 0,
\end{equation}
where momenta are defined by $p_t = \partial_t S$, $p_i = \partial_i
S$, and $p^2 = p_i p^i = g^{ij} p_i p_j$. In order to recover the
dispersion relation in general relativity, we can take $\Lambda_0 =
m^2 c^2$ and $\Lambda_1 =1$, and assume that $\Lambda_n$'s are very
small for $n\ge 2$. For convenience, let us define $x^\mu = (ct,
x^i)$, and the Boltzman constant is set to be $k_B = 1$. For the
spherically symmetric background of $N^i=0$, the dispersion relation
(\ref{WKB}) can be written as
\begin{equation}
  \label{mom:vacuum}
  p_0 p^0 + \sum_{n=1}^z \Lambda_n (p_i p^i)^n = -m^2 c^2,
\end{equation}
where $p_\mu$ is the conjugate momentum to $x^\mu$. The constants
$\Lambda_n$ have a scaling dimension of $(2n - 2)$. Note that
Eq.~(\ref{mom:vacuum}) can be reduced to $p_0 p^0 + p_i p^i = -m^2
c^2$ for the relativistic limit of $z=1$.

Now, we choose the constants $\Lambda_n$ as $\Lambda_n = \ell_{\rm
  P}^{2(z-1)} \delta_{nz}$ for $n \ge 1$ to obtain a reasonable
entropy where $\ell_{\rm P}$ is the Plank length given by $\ell_{\rm
  P} = \sqrt{\hbar G_N /c^3}$.  As a result, the dispersion relation
(\ref{mom:vacuum}) for the scalar field with mass $m$ can be simply
reduced to
\begin{equation}
  \label{mom:scalar}
  p_0 p^0 + \ell_{\rm P}^{2(z-1)} (p_i p^i)^z = -m^2 c^2.
\end{equation}
We will consider this reduced relation of the highest momentum case
for a scalar field on spherically symmetric black hole background.

\section{Entropy in reduced dispersion relation}
\label{sec:entropy:HL}

We now consider a spherically symmetric black hole whose line element
can be written as
\begin{equation}
  \label{metric:bh}
  ds^2 = -f \tilde{N}^2  c^2 dt^2 + \frac{dr^2}{f} + r^2 (d\theta^2 +
  \sin^2\theta\, d\phi^2),
\end{equation}
where $f=f(r)$ and $\tilde{N} = \tilde{N}(r)$, and the horizon $r_H$
of the black hole is defined by $g^{rr}|_{r_H} = f(r_H) = 0$. With the
help of conjugate pairs of $x^\mu = (ct, r, \theta, \phi)$ and $p_\mu
= (-\omega/c, p_r, p_\theta, p_\phi)$, the dispersion
relation~(\ref{mom:scalar}) becomes
\begin{equation}
  \label{mom:metric}
  \ell_{\rm P}^{2(z-1)} \left(f p_r^2 + \frac{p_\theta^2}{r^2} +
    \frac{p_\phi^2}{r^2\sin^2\theta} \right)^z =
  \frac{\omega^2}{f \tilde{N}^2c^2} - m^2 c^2.
\end{equation}

Let us consider a spherical box specified by $r_H + \epsilon$ to $r_H
+ \epsilon + \delta $ where $\epsilon$ plays a role of UV cutoff in
the conventional brick wall method. It will be shown that it is
nonnecessary because the UV divergent behavior of free energy can be
improved. Next, the number of quantum states with the energy less than
$\omega$ is calculated as
\begin{equation}
  \label{def:n}
  n(\omega) = \frac{1}{(2\pi)^3} \int dr d\theta d\phi dp_r dp_\theta
  dp_\phi = \frac{1}{(2\pi)^3} \int dr d\theta d\phi \times V_p,
\end{equation}
where $V_p$ is the $z$-dimensional volume of momentum space satisfying
Eq.~(\ref{mom:metric}), which is explicitly
\begin{equation}
  \label{n:UV}
  n(\omega) = \frac{2 }{3\pi \ell_{\rm P}^{\frac{3}{z}(z-1)}  } \int dr \frac{r^2}{\sqrt{f}}
  \left( \frac{\omega^2}{f \tilde{N}^2 c^2} - m^2 c^2 \right)^{\frac{3}{2z}},
\end{equation}
near the horizon where $\omega$ is the energy of a scalar field with
the range of $\omega \ge m c^2 \tilde{N}\sqrt{f}$.  For $z=1$, it
recovers the standard form of number of the quantum states,
\begin{equation}
 \label{n:IR}
 n(\omega) = \frac{2}{3\pi} \int dr \frac{r^2}{\sqrt{f}}
  \left( \frac{\omega^2}{f \tilde{N}^2 c^2} - m^2 c^2 \right)^{\frac{3}{2}}.
\end{equation}
Then, the free energy is given by
\begin{equation}
  \label{F}
  F_{(z)} = - \int_{\omega_0}^\infty d\omega \frac{n(\omega)}{e^{\beta\omega}-1},
\end{equation}
where $\omega_0 = m c^2 \tilde{N} \sqrt{f}$ and $\beta^{-1}$ is a
inverse temperature defined by $\beta^{-1}= \kappa_H/ (2\pi c)$, and
$\kappa_H = \frac12 c^2 \tilde{N} f'|_{r_H}$ is a surface gravity. So,
the entropy can be written as
\begin{equation}
  \label{S:n}
  S_{(z)} = \beta^2 \frac{\partial F_{(z)}}{\partial\beta} = \beta^2
  \int_{\omega_0}^\infty \, d\omega \frac{\omega\, n(\omega)}{4
    \sinh^2 \frac12\beta\omega}.
\end{equation}
For the sake of convenience, $\omega$ is replaced by $x = \frac12
\beta \omega$. Then, it can be written as
\begin{equation}
  \label{S:n:x}
  S_{(z)} = \int_{x_0}^\infty \, dx \frac{x\, n(2x/\beta)}{\sinh^2 x},
\end{equation}
where $x_0 = \frac12\beta m c^2 \tilde{N} \sqrt{f}$, which goes to
zero near the horizon. Plugging Eq.~(\ref{n:UV}) into
Eq.~(\ref{S:n:x}), the entropy becomes
\begin{equation}
  \label{S:x}
  S_{(z)} = \frac{2 }{3\pi} \int dr\, \frac{r^2}{\sqrt{f}}
  \int_{x_0}^\infty dx \, \frac{x J(x) }{\sinh^2 x},
\end{equation}
where
\begin{equation}
  \label{I:def}
  J(x) \equiv \left\{
    \begin{array}{ll}
      \ell_{\rm P}^{\frac{3}{z}(1-z)} \left(
        \frac{4x^2}{\beta^2 f \tilde{N}^2 c^2} - m^2 c^2 \right)^{\frac{3}{2z}}, &
      \textrm{near the horizon} \\
      \left(
        \frac{4x^2}{\beta^2 f \tilde{N}^2 c^2} - m^2 c^2 \right)^{\frac{3}{2}}, & \textrm{for
        $r \gg r_H$}
    \end{array} \right.
\end{equation}
Note that for $r \gg r_H$, as expected, the integral is proportional
to the volume of space as long as the metric function $f$ approaches
nonzero constant at infinity.

On the other hand, we are concerned about the entropy near the
horizon, which is given by
\begin{equation}
  \label{S:r}
  S_{(z)} = \frac{2}{3\pi \ell_{\rm P}^{\frac{3}{z}(z-1)}  }
  \int dr \, \frac{r^2}{f^{(1+3/z)/2}} \left( \frac{2}{\beta c \tilde{N}} \right)^{\frac{3}{z}} \alpha(z)
\end{equation}
where $\alpha(z) \equiv \int_0^\infty dx \, \frac{x^{1+3/z}}{\sinh^2
  x}$.  For some of $z$, we can find $\alpha(1) = \pi^4/30$,
$\alpha(2) \approx 1.5762$, $\alpha(3) = \pi^2/6$, and $\alpha(4)
\approx 1.8766$. Now, in the near horizon limit, the function $f(x)$
can be expanded as $f(r) = \frac{2\kappa_H}{\tilde{N}_H c^2} (r-r_H) +
O(r-r_H)^2$ with $\tilde{N}_H \equiv \tilde{N}(r_H)$, so that one can
take the first-order approximation of $\kappa_H \ne 0$ for nonextremal
black holes. Therefore, the entropy (\ref{S:r}) explicitly calculated
as
\begin{eqnarray}
  S_{(3)} &=& \frac{A}{4} \cdot \frac{c}{9\beta \ell_{\rm P}^2 \kappa_H} \cdot \ln
  \left( 1+ \frac{\delta}{\epsilon} \right), \qquad \mathrm{for}\ z=3
  \label{S:z:3:layer} \\
  S_{(z)} &=& \frac{A}{4} \cdot \frac{4z\alpha(z)}{3\pi^2(z-3) \ell_{\rm
      P}^{\frac{3}{z}(z-1)}  } \left(\frac{2}{\beta c \tilde{N}_H}
  \right)^{\frac{3}{z}}
  \left(\frac{\tilde{N}_H c^2}{2\kappa_H} \right)^{\frac12 +
    \frac{3}{2z}} \left[ (\epsilon + \delta)^{\frac12- \frac{3}{2z}} -
    \epsilon^{\frac12 - \frac{3}{2z}} \right], \ \mathrm{for}\  z \ne
  3.  \label{S:z:layer}
\end{eqnarray}
Note that there is no continuous limit at $z=3$. Now, defining proper
lengths for $\epsilon$ and $\delta$, respectively
\begin{eqnarray}
  \bar\epsilon &\equiv& \int_{r_H}^{r_H+\epsilon} dr \sqrt{g_{rr}}  =
  \sqrt{\frac{2c^2 \tilde{N}_H
      \epsilon}{\kappa_H}}, \label{properlength:epsilon} \\
  \bar\delta &\equiv& \int_{r_H+\epsilon}^{r_H+\epsilon+\delta} dr \sqrt{g_{rr}}  =
  \sqrt{\frac{2c^2 \tilde{N}_H}{\kappa_H}} (\sqrt{\epsilon+\delta} -
  \sqrt{\epsilon}), \label{properlength:delta}
\end{eqnarray}
where $\epsilon = \frac12 \kappa_H \bar\epsilon^2 /(\tilde{N}_H c^2)$
and $\epsilon+\delta = \frac12 \kappa_H (\bar\epsilon +
\bar\delta)^2/(\tilde{N}_H c^2)$, the entropy can be written as
\begin{eqnarray}
  S_{(3)} &=& \frac{A}{4} \frac{1}{9\pi \ell_{\rm P}^2} \ln
  \left(1+\frac{\bar\delta}{\bar\epsilon} \right) \qquad \mathrm{for}
  \ z=3 \label{S:z:3:final}, \\
  S_{(z)} &=& \frac{A}{4} \cdot \frac{2z \alpha(z)}{3\pi^{2+\frac{3}{z}}
    (z-3) \ell_{\rm P}^{\frac{3}{z}(z-1)}} \left[(\bar\epsilon  +
    \bar\delta)^{1-\frac{3}{z}} - \bar\epsilon^{1-\frac{3}{z}} \right] \ \mathrm{for}\
  z \ne 3. \label{S:z:final}
\end{eqnarray}
It is interesting to note that for the case of $z >3$, the
entropy~(\ref{S:z:final}) is finite
\begin{equation}
  \label{S:final:zge4}
  S_{(z)} = \frac{A}{4} \cdot \frac{2z \alpha(z)}{3\pi^{2+\frac{3}{z}} (z-3) \ell_{\rm P}^{\frac{3}{z}(z-1)} }
  \bar\delta^{1-\frac{3}{z}},
\end{equation}
even in spite of the absence of the UV cutoff i.e., $\bar\epsilon \to
0$. In other words, for the case of $z \le 3 $, the UV cutoff is
necessary to get some finite results.  Recovering the dimension except
the Boltzman constant $k_B$, the entropy (\ref{S:final:zge4}) is
written as
\begin{equation}
  \label{S:final:z:dim}
  S_{(z)} = \frac{c^3 A}{4\hbar G_N} \cdot \frac{2z\alpha(z) \,
    \bar\delta^{1-\frac{3}{z}}}{3\pi^{2+\frac{3}{z}} (z-3) \ell_{\rm P}^{\frac{z-3}{z}}}.
\end{equation}
Then, Eq.~(\ref{S:final:z:dim}) is compatible with the
Bekenstein-Hawking entropy given by
\begin{equation}
S_{\rm BH} = \frac{c^3 A}{4\hbar G_N}
\end{equation}
as long as we identify the size of box as
\begin{equation}
  \label{thickness}
  \bar\delta = \ell_{\rm P} \left[\frac{3(z-3)
  \pi^{2+\frac{3}{z}}}{2z\alpha(z)} \right]^{\frac{z}{z-3}},
\end{equation}
for $z >3$. It depends on the scale parameter $z$, however, it is
independent of the black hole hairs.

There are some special limits to be mentioned.  As for the marginal
case of $z=3$, recovering dimensions, the entropy becomes
\begin{equation}
  \label{S:final:thin:3:dim}
  S_{(3)} = \frac{c^3 A}{4\hbar G_N} \cdot \frac{1}{9\pi} \ln \left( 1 +
    \frac{\bar\delta}{\bar\epsilon} \right).
\end{equation}
Note that Eq.~(\ref{S:final:thin:3:dim}) also agrees with the
Bekenstein-Hawking entropy assuming $\bar\delta/\bar\epsilon =
e^{9\pi} - 1$. In this case, the UV cutoff is needed. On the other
hand, for the limit of $z=1$ which corresponds to the (thin layered)
brick wall method, the well-known cutoff parameter can be obtained, $
S_{(1)} = c^3A/(4\hbar G_N) \cdot \delta \ell_{\rm P}^2/[90\beta
\epsilon(\epsilon+\delta)]$~\cite{Liu:2001ra}.  In these respects,
excitations of the Lifshitz scalar field coupled to the gravity
contributes to the finite entropy near horizon limit without UV cutoff
for certain scaling parameter.

\section{Entropy from (partially) modified dispersion relation}
\label{sec:z:even}

We are going to devote this section to clarify some issues related to
the modified dispersion relation~(\ref{mom:vacuum}). In the course of
calculations, we have considered just the highest power of
nonisotropic exponent in the dispersion relation for simplicity, such
as $\Lambda_n$'s are chosen as $\Lambda_n =\ell_{\rm P}^{2(n-1)}$ for
$n \ge 1$.  The justification for this is needed from a general point
of view because all terms in the modified dispersion relation may
contribute to the final entropy. For instance, the large infrared (IR)
cutoff $\bar\delta$ may require all contributions of terms. To answer
to this question, instead of analytic results, we want to present some
numerical simulations in order to show how much the previous results
change for $z=2$ and $z=4$ when we take a partially modified
dispersion relation which is a sort of reduced relation.

Now, for the solvability, we take a partially modified dispersion
relation as follows:
\begin{equation}
  \label{dispersion:z:even}
  p_0 p^0 + \Lambda_{\frac{z}{2}} (p_i p^i)^{\frac{z}{2}} + \Lambda_z (p_i p^i)^z = - m^2 c^2,
\end{equation}
where $z$ is even. The full dispersion relation, for instance, for the
case of $z=2$ recovers as $ p_0 p^0 + \Lambda_1 p_i p^i + \Lambda_2
(p_i p^i)^2 = - m^2 c^2$.  In spherically symmetric black holes
described by the line element (\ref{metric:bh}), the dispersion
relation (\ref{dispersion:z:even}) becomes
\begin{eqnarray}
  p_i p^i &=& f p_r^2 + \frac{p_\theta^2}{r^2} + \frac{p_\phi^2}{r^2\sin^2\theta} \nonumber \\
  &=& \frac{1}{(2\Lambda_z)^{\frac2z}} \left[ -\Lambda_{\frac{z}{2}} +
    \sqrt{\Lambda_{\frac{z}{2}} + 4 \Lambda_z \left(\frac{\omega^2}{f
          \tilde{N}^2 c^2} - m^2 c^2 \right)} \right]^{\frac2z}. \label{p:z:even}
\end{eqnarray}
Of course, for $\Lambda_z = 0$ and $z=2$, the number of quantum states
can be reduced to the relativistic limit,
\begin{equation}
  \label{p:relativistic:z:even}
  p_i p^i = \frac{1}{\Lambda_1} \left( \frac{\omega^2}{f \tilde{N}^2 c^2} - m^2 c^2
  \right). 
\end{equation}
Next, from Eq.~(\ref{p:z:even}), the number of quantum states
with the energy less than $\omega$ can be written as
\begin{equation}
  n(\omega) = \frac{2}{3\pi (2\Lambda_z)^{\frac3z}} \int dr
  \frac{r^2}{\sqrt{f}} \left[ -\Lambda_{\frac{z}{2}} +
    \sqrt{\Lambda_{\frac{z}{2}} ^2 + 4\Lambda_z
      \left(\frac{\omega^2}{f \tilde{N}^2 c^2} - m^2 c^2 \right)}
  \right]^{\frac3z}, \label{n:z:even} 
\end{equation}
where the energy should satisfy $\omega \ge \omega_0 \equiv mc^2
\tilde{N} \sqrt{f}$.  Substituting Eq.~(\ref{n:z:even}) into
Eq.~(\ref{S:n:x}), the entropy is written as
\begin{equation}
  S = \frac{2}{3\pi (2\Lambda_z)^{\frac3z}} \int_{x_0}^\infty dx
  \frac{x}{\sinh^2 x} \int dr \frac{r^2}{\sqrt{f}} \left[ -\Lambda_{\frac{z}{2}}  +
    \sqrt{\Lambda_{\frac{z}{2}}^2 + \frac{16 \Lambda_4}{\beta^2 c^2 \tilde{N}^2 f}
      \left( x^2 - x_0^2 \right)} \right]^{\frac3z}, \label{S:z:even}
\end{equation}
where $x$ and $x_0$ are defined by $x \equiv \frac12 \beta \omega$ and
$x_0 = \frac12 \beta m c^2 \tilde{N} \sqrt{f}$. For the thin layer
with the range from $r+\epsilon$ to $r+\epsilon + \delta$ near the
horizon, it can be written as
\begin{equation}
  S = \frac{2c}{3\pi (2\Lambda_4)^{\frac3z}}
  \sqrt{\frac{\tilde{N}_H}{\kappa_H}} \int_{r_H + \epsilon}^{r_H +
    \epsilon + \delta} \hspace{-5mm} dr\, \frac{r_H^2}{\sqrt{r -
      r_H}} \int_{x_0}^\infty \! dx\, \frac{x}{\sinh^2 x} \left[
    -\Lambda_{\frac{z}{2}} + \sqrt{\Lambda_{\frac{z}{2}}^2 + \frac{8
        \Lambda_z}{\beta^2 \tilde{N}_H \kappa_H} \frac{x^2 -
        x_0^2}{r - r_H}} \right]^{\frac3z}. \label{S:thin:z:even}
\end{equation}

\begin{figure}[pbt]
  \centering
  \subfigure[]{\label{fig:S2all}\includegraphics[width=0.45\textwidth]{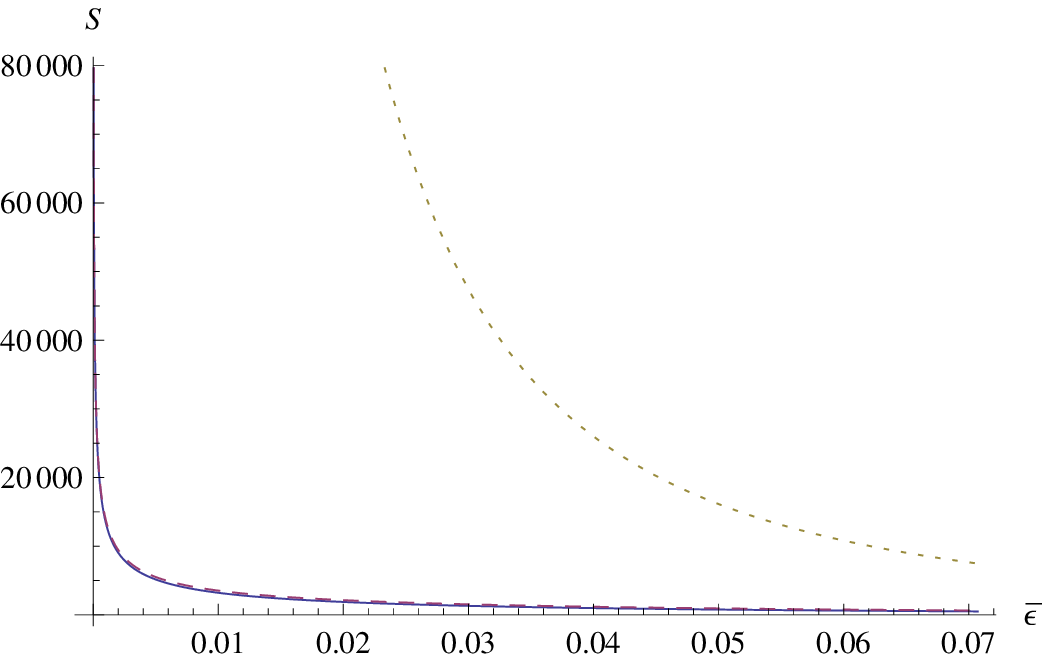}} 
  \hfill
  \subfigure[]{\label{fig:S2part}\includegraphics[width=0.45\textwidth]{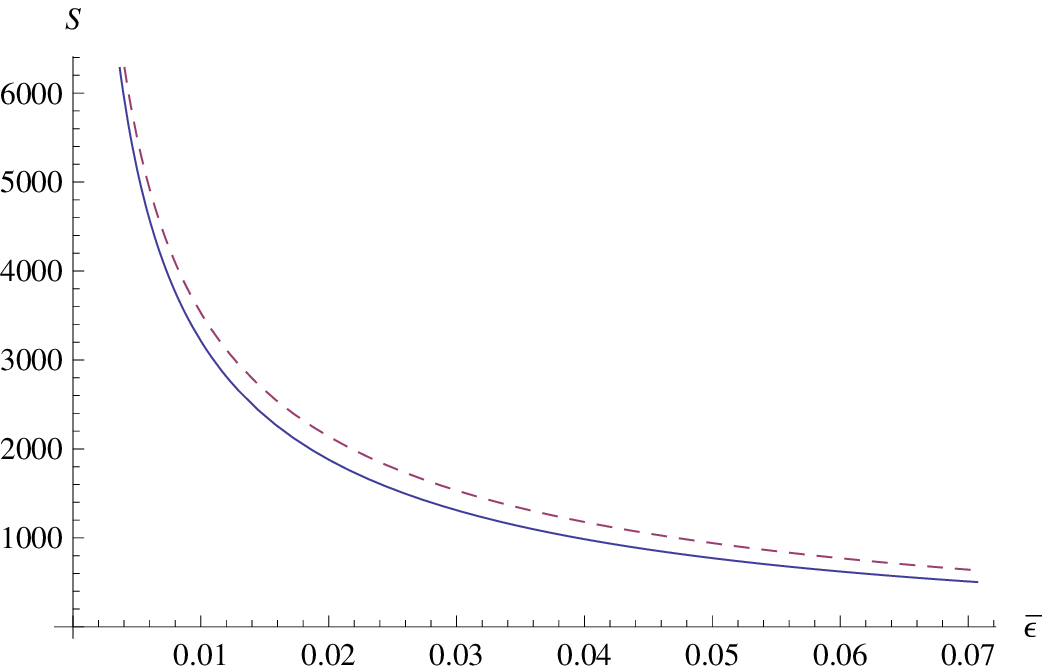}}
  \caption{It shows the entropies as a function of the cutoff
    parameter $\bar\epsilon$ for the case of $z=2$. The dotted,
    dashed, and solid lines correspond to the cases of $R_2 =
    (\Lambda_1, \Lambda_2) = (1, 0)$, $H_2 = (\Lambda_1, \Lambda_2) =
    (0, \ell^2)$, and $F_2=(\Lambda_1, \Lambda_2) = (1, \ell^2)$,
    respectively, where $\ell = 0.01$.  The variables in
    Eq.~(\ref{S:thin:z:even}) have been simply chosen as $\delta =
    0.05$, $r_H = \kappa_H = 2$, and $\tilde{N}_H = m = c = 1$.}
  \label{fig:S2}
\end{figure}

\begin{figure}[pbt]
  \centering
  \includegraphics[width=0.5\textwidth]{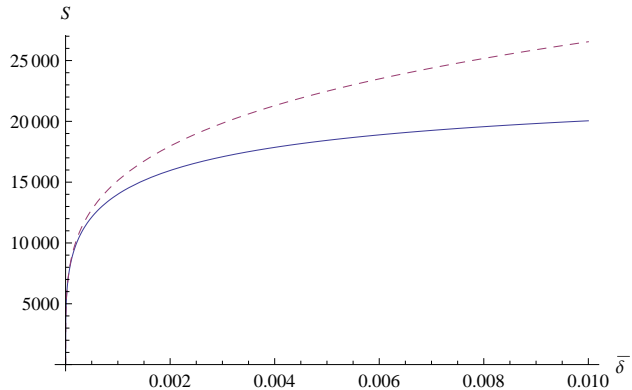}
  \caption{It shows the entropies as the function of the cutoff
    parameter $\bar\delta$ for the case of $z=4$. The dashed and solid
    lines correspond to the cases of $H_4 = (\Lambda_2, \Lambda_4) =
    (0, \ell^6)$ and $F_4 = (\Lambda_2, \Lambda_4) = (\ell^2,
    \ell^6)$, respectively, where $\ell = 0.01$.  The variables in
    Eq.~(\ref{S:thin:z:even}) have been chosen as $\epsilon=0$, $r_H =
    \kappa_H = 2$, and $\tilde{N}_H = m = c = 1$.}
  \label{fig:S4}
\end{figure}

Actually, it is not easy to get analytic results, so that we plot
entropies with respect to the proper lengths $\bar\epsilon$ for the
case of $z=2$ and $\bar\delta$ for the case of $z=4$, which are shown
in Figs.~\ref{fig:S2} and~\ref{fig:S4}, respectively. The coefficients
of momenta in the dispersion relation~(\ref{dispersion:z:even}) have
been chosen as $\Lambda_n = \ell^{2(n-1)}$ for $n=z, z/2$ for the sake
of comparison with the results obtained in
section~\ref{sec:entropy:HL}.

For $z=2$, there are largely three cases, a relativistic limit, $R_2
\equiv (\Lambda_1, \Lambda_2) = (1, 0)$, a highest momentum
consideration, $H_2 \equiv (\Lambda_1, \Lambda_2)= (0, \ell^2)$, and a
full consideration, $F_2 \equiv (\Lambda_1, \Lambda_2) = (1, \ell^2)$.
As shown in Fig.~\ref{fig:S2all}, much smaller cutoffs are required
compared to that of the relativistic case for the same value of the
entropy, so that the curve for $R_2$ case lies far above these two
curves of $H_2$ and $F_2$.  This fact has been discussed using a
modified dispersion which is similar to the full dispersion relation
($F_2$) of $z=2$~\cite{Corley:1996ar,Jacobson:2007jx}, in particular,
it is interesting to see that for the black hole entropy, the same
brick wall lies at a much smaller proper distance in the free-fall
time slice compared to the conventional brick wall cutoff. While the
conventional brick wall cuts off all modes at the same location, they
cutoff all modes at the same momentum. By the way, from
Fig.~\ref{fig:S2part}, we can see that the cutoff in $F_2$ case is
slightly smaller than that of $H_2$ case.  As a result, the full
consideration of the dispersion relation is very close to the highest
momentum consideration in comparison with the relativistic case, even
though the former case gives a slightly smaller cutoff.

On the other hand, in order to investigate the role of the lower
momentum contribution for $z >3$, we study the dispersion
relation~(\ref{dispersion:z:even}) for the specific case of $z=4$.
The entropy curves with respect to the proper IR cutoff $\bar\delta$
are plotted in Fig.~\ref{fig:S4} for the highest power of momentum
consideration, $H_4 \equiv (\Lambda_2, \Lambda_4) = (0, \ell^6)$ and
the full consideration, $F_4 \equiv (\Lambda_2, \Lambda_4) = (\ell^2,
\ell^6)$, respectively. It can be shown that in a very short distance
compared to the given length scale of $\ell$, the entropy profiles are
almost coincident, while they generate a little difference as the
cutoff is getting large. So, if $\bar\delta$ is much smaller than
$\ell$, then low momentum contributions can be neglected, whereas
their contributions cannot be ignored when $\bar\delta$ and $\ell$ are
at the same order of scale.  It means that we have to discard the area
law of the entropy at $\bar\delta \approx \ell$.  Therefore, the full
modified dispersion relation will give the smaller entropy, which is
not compatible with the area law.  Of course, this conclusion is more
or less restrictive, so we hope this issue will be discussed more
generally elsewhere.

\section{Discussions}
\label{sec:discussion}

We have studied the statistical entropy of spherical symmetric black
holes using the brick wall method in HL gravity. The crucial
difference from the conventional brick wall method is that the scalar
field satisfying FPD called Lifshitz scalar gives the area law of the
finite entropy without UV cutoff for $z >3 $ corresponding to the
super-renormalizable sector of HL gravity as long as the length of
thin wall is identified with a certain value depending on the scale
parameter.

This result is reminiscent of the entropy calculation in the brick
wall method using GUP, $\Delta x \Delta p \ge \hbar +
\frac{\lambda}{\hbar} (\Delta p)^2$, and there exists a minimal
length, $\Delta x_{\rm{min}} = 2 \sqrt{\lambda}$ \cite{Garay:1994en,
  Chang:2001bm, Yoon:2007aj}.  Similar to the present modes counting
between the horizon and $\bar{\delta}$, it happens between just
outside horizon and the minimal length, $\Delta x_{\rm{min}}$ without
any UV cutoff in GUP regime.  For instance, in a spherical symmetric
black hole based on GUP, the number of quantum states in this minimal
length is obtained as $ n(\omega) = \frac{2}{3\pi} \int dr \frac{r^2
  (\omega^2/f - \mu^2)^{3/2}}{\sqrt{f}[1 + \lambda (\omega^2/f -
  \mu^2)]^3}$.  Following the same procedure with the previous
section~\ref{sec:entropy:HL}, the entropy is calculated as $ S =
\frac{c^3 A}{4\hbar G_N} \cdot \frac{\xi \ell_{\rm P}^2}{\lambda}$,
where $\xi \equiv \frac13 [\frac{4}{\pi}\zeta(3) - \frac{25}{8\pi} -
\frac{\pi}{6}]$.  To satisfy the area law of entropy for this black
hole, $\lambda$ is required to be the same with $\xi \ell_{\rm
  P}^2$. In comparison with HL gravity, the thickness of thin wall of
$\bar{\delta}$ can be identified with the minimal length in GUP
$\bar{\delta} = 2\sqrt{\lambda}$. Then, using Eq.~(\ref{thickness}),
it amounts the highly super-renormalizable case of HL gravity
$z\approx 342$.  It implies that the Lifshitz scalar field may play a
role of the usual scalar field along with GUP at least in the brick
wall regime. Unfortunately, it is unclear why the scaling parameter
should be so large if we try to match the minimal length in GUP and
$\bar{\delta}(z)$.

%%%%%% PRD
\begin{acknowledgments}
  We would like to thank E.\ Son for exciting discussions.  W. KIM was
  supported by the Basic Science Research Program through the National
  Research Foundation of Korea(NRF) funded by the Ministry of
  Education, Science and Technology(2010-0008359), and the Special
  Research Grant of Sogang University, 200911044.  M. Eune was
  supported by the National Research Foundation of Korea Grant funded
  by the Korean Government [NRF-2009-351-C00109].
\end{acknowledgments}

%%%%%%%%%%%%%%%%%%%%%%%%%%%%%%%%%%%%%%%%%%%%%%%%%%%%%%%%%%%%%
%%%%%%%%%%%%%%%             References       %%%%%%%%%%%%%%%%
%%%%%%%%%%%%%%%%%%%%%%%%%%%%%%%%%%%%%%%%%%%%%%%%%%%%%%%%%%%%%

% Create the reference section using BibTeX:
%\bibliography{basename of .bib file}

\begin{thebibliography}{99}

\bibitem{Horava:2008jf}
  P.~Horava,
  %``Quantum Criticality and Yang-Mills Gauge Theory,''
  arXiv:0811.2217 [hep-th];
  %%CITATION = ARXIV:0811.2217;%%
%\bibitem{Horava:2008ih}
  %P.~Horava,
  %``Membranes at Quantum Criticality,''
  JHEP {\bf 0903}, 020 (2009)
  [arXiv:0812.4287 [hep-th]].
  %%CITATION = JHEPA,0903,020;%%

\bibitem{Horava:2009uw}
  P.~Horava,
  %``Quantum Gravity at a Lifshitz Point,''
  Phys.\ Rev.\  D {\bf 79}, 084008 (2009)
  [arXiv:0901.3775 [hep-th]].
  %%CITATION = PHRVA,D79,084008;%%

\bibitem{Blas:2009qj}
  D.~Blas, O.~Pujolas and S.~Sibiryakov,
  %``Consistent Extension Of Horava Gravity,''
  Phys.\ Rev.\ Lett.\  {\bf 104}, 181302 (2010)
  [arXiv:0909.3525 [hep-th]];
  %%CITATION = PRLTA,104,181302;%%
%\bibitem{Blas:2009ck}
  %D.~Blas, O.~Pujolas and S.~Sibiryakov,
  %``Comment on `Strong coupling in extended Horava-Lifshitz gravity',''
  Phys.\ Lett.\  B {\bf 688}, 350 (2010)
  [arXiv:0912.0550 [hep-th]].
  %%CITATION = PHLTA,B688,350;%%

\bibitem{Wang:2009azb}
  A.~Wang, D.~Wands and R.~Maartens,
  %``Scalar field perturbations in Horava-Lifshitz cosmology,''
  JCAP {\bf 1003}, 013 (2010)
  [arXiv:0909.5167 [hep-th]].
  %%CITATION = JCAPA,1003,013;%%

\bibitem{Lu:2009em}
  H.~Lu, J.~Mei and C.~N.~Pope,
  %``Solutions to Horava Gravity,''
  Phys.\ Rev.\ Lett.\  {\bf 103}, 091301 (2009)
  [arXiv:0904.1595 [hep-th]].
  %%CITATION = PRLTA,103,091301;%%

\bibitem{Cai:2009pe}
  R.~G.~Cai, L.~M.~Cao and N.~Ohta,
  %``Topological Black Holes in Horava-Lifshitz Gravity,''
  Phys.\ Rev.\  D {\bf 80}, 024003 (2009)
  [arXiv:0904.3670 [hep-th]].
  %%CITATION = PHRVA,D80,024003;%%

\bibitem{Cai:2009ar}
  R.~G.~Cai, Y.~Liu and Y.~W.~Sun,
  %``On the z=4 Horava-Lifshitz Gravity,''
  JHEP {\bf 0906}, 010 (2009)
  [arXiv:0904.4104 [hep-th]].
  %%CITATION = JHEPA,0906,010;%%

\bibitem{Colgain:2009fe}
  E.~O.~Colgain and H.~Yavartanoo,
  %``Dyonic solution of Horava-Lifshitz Gravity,''
  JHEP {\bf 0908}, 021 (2009)
  [arXiv:0904.4357 [hep-th]].
  %%CITATION = JHEPA,0908,021;%%

\bibitem{Kehagias:2009is}
  A.~Kehagias and K.~Sfetsos,
  %``The black hole and FRW geometries of non-relativistic gravity,''
  Phys.\ Lett.\  B {\bf 678}, 123 (2009)
  [arXiv:0905.0477 [hep-th]].
  %%CITATION = PHLTA,B678,123;%%

\bibitem{Myung:2009va}
  Y.~S.~Myung,
   %``Thermodynamics of black holes in the deformed Ho\v{r}ava-Lifshitz
  %gravity,''
  Phys.\ Lett.\  B {\bf 678}, 127 (2009)
  [arXiv:0905.0957 [hep-th]];
  %%CITATION = PHLTA,B678,127;%%
%\bibitem{Ghodsi:2009zi}
  A.~Ghodsi and E.~Hatefi,
  %``Extremal rotating solutions in Horava Gravity,''
  Phys.\ Rev.\  D {\bf 81}, 044016 (2010)
  [arXiv:0906.1237 [hep-th]];
  %%CITATION = PHRVA,D81,044016;%%
%\bibitem{Greenwald:2009kp}
  J.~Greenwald, A.~Papazoglou and A.~Wang,
  %``Black holes and stars in Horava-Lifshitz theory with projectability
  %condition,''
  Phys.\ Rev.\  D {\bf 81}, 084046 (2010)
  [arXiv:0912.0011 [hep-th]].
  %%CITATION = PHRVA,D81,084046;%%

\bibitem{Cai:2010ud}
  R.~G.~Cai and A.~Wang,
  %``Singularities in Horava-Lifshitz theory,''
  Phys.\ Lett.\  B {\bf 686}, 166 (2010)
  [arXiv:1001.0155 [hep-th]];
  %%CITATION = PHLTA,B686,166;%%
%\bibitem{Ding:2010kr}
  C.~Ding, S.~Chen, J.~Jing,
  %``Analytical expression for the greybody factor and dynamic
  %evolution for the scalar field in the Ho\v{r}ava-Lifshitz black
  %hole,''
  [arXiv:1004.4440 [gr-qc]].

\bibitem{Koutsoumbas:2010pt}
  G.~Koutsoumbas, E.~Papantonopoulos, P.~Pasipoularides and M.~Tsoukalas,
  %``Black Hole Solutions in 5D Horava-Lifshitz Gravity,''
  Phys.\ Rev.\  D {\bf 81} (2010) 124014
  [arXiv:1004.2289 [hep-th]];
  %%CITATION = PHRVA,D81,124014;%%
%\bibitem{Koutsoumbas:2010yw}
  G.~Koutsoumbas and P.~Pasipoularides,
  %``Black hole solutions in Horava-Lifshitz Gravity with cubic terms,''
  arXiv:1006.3199 [hep-th].
  %% CITATION = ARXIV:1006.3199;%%

\bibitem{Kiritsis:2009sh}
  E.~Kiritsis and G.~Kofinas,
  %``Horava-Lifshitz Cosmology,''
  Nucl.\ Phys.\  B {\bf 821}, 467 (2009)
  [arXiv:0904.1334 [hep-th]].
  %%CITATION = NUPHA,B821,467;%%

\bibitem{Brandenberger:2009yt}
  R.~Brandenberger,
  %``Matter Bounce in Horava-Lifshitz Cosmology,''
  Phys.\ Rev.\  D {\bf 80}, 043516 (2009)
  [arXiv:0904.2835 [hep-th]];
  %%CITATION = PHRVA,D80,043516;%%
%\bibitem{Mukohyama:2009zs}
  S.~Mukohyama, K.~Nakayama, F.~Takahashi and S.~Yokoyama,
  %``Phenomenological Aspects of Horava-Lifshitz Cosmology,''
  Phys.\ Lett.\  B {\bf 679}, 6 (2009)
  [arXiv:0905.0055 [hep-th]];
  %%CITATION = PHLTA,B679,6;%%
%\bibitem{Cai:2009in}
  Y.~F.~Cai and E.~N.~Saridakis,
  %``Non-singular cosmology in a model of non-relativistic gravity,''
  JCAP {\bf 0910}, 020 (2009)
  [arXiv:0906.1789 [hep-th]].
  %%CITATION = JCAPA,0910,020;%%

\bibitem{Kiritsis:2009vz}
  E.~Kiritsis,
  %``Spherically symmetric solutions in modified Horava-Lifshitz gravity,''
  Phys.\ Rev.\  D {\bf 81}, 044009 (2010)
  [arXiv:0911.3164 [hep-th]].
  %%CITATION = PHRVA,D81,044009;%%

\bibitem{Majhi:2009xh}
  B.~R.~Majhi,
  %``Hawking radiation and black hole spectroscopy in Horava-Lifshitz gravity,''
  Phys.\ Lett.\  B {\bf 686}, 49 (2010)
  [arXiv:0911.3239 [hep-th]];
  %%CITATION = PHLTA,B686,49;%%
%\bibitem{Setare:2010wt}
  M.~R.~Setare and M.~Jamil,
  %``Holographic dark energy with varying gravitational constant in
  %Horava-Lifshitz cosmology,''
  JCAP {\bf 1002}, 010 (2010)
  [arXiv:1001.1251 [hep-th]];
  %%CITATION = JCAPA,1002,010;%%
%\bibitem{Jamil:2010di}
  M.~Jamil, E.~N.~Saridakis and M.~R.~Setare,
  %``The generalized second law of thermodynamics in Horava-Lifshitz
  %cosmology,''
  arXiv:1003.0876 [hep-th].
  %%CITATION = ARXIV:1003.0876;%%

\bibitem{Son:2010qh}
  E.~J.~Son and W.~Kim,
  %``Smooth cosmological phase transition in the Horava-Lifshitz gravity,''
  JCAP {\bf 1006}, 025 (2010)
  [arXiv:1003.3055 [hep-th]];
  %%CITATION = JCAPA,1006,025;%%
%\bibitem{Dutta:2010jh}
  S.~Dutta and E.~N.~Saridakis,
  %``Overall observational constraints on the running parameter \lambda of
  %Horava-Lifshitz gravity,''
  JCAP {\bf 1005}, 013 (2010)
  [arXiv:1002.3373 [hep-th]];
  %%CITATION = JCAPA,1005,013;%%
%\bibitem{Jamil:2010vr}
  M.~Jamil and E.~N.~Saridakis,
  %``New agegraphic dark energy in Horava-Lifshitz cosmology,''
  JCAP {\bf 1007}, 028 (2010)
  [arXiv:1003.5637 [physics.gen-ph]];
  %%CITATION = JCAPA,1007,028;%%
%\bibitem{Ali:2010sv}
  A.~Ali, S.~Dutta, E.~N.~Saridakis and A.~A.~Sen,
  %``Horava-Lifshitz cosmology with generalized Chaplygin gas,''
  arXiv:1004.2474 [astro-ph.CO].
  %%CITATION = ARXIV:1004.2474;%%

\bibitem{Romero:2009qs}
  J.~M.~Romero, V.~Cuesta, J.~A.~Garcia and J.~D.~Vergara,
  %``Conformal Anisotropic Mechanics,''
  Phys.\ Rev.\ D {\bf 81}, 065013 (2010)
  [arXiv:0909.3540 [hep-th]].
  %%CITATION = ARXIV:0909.3540;%%

\bibitem{Rama:2009xc}
  S.~K.~Rama,
  %``Particle Motion with Ho\v{r}ava -- Lifshitz type Dispersion Relations,''
  arXiv:0910.0411 [hep-th].
  %%CITATION = ARXIV:0910.0411;%%

\bibitem{Corley:1996ar}
  S.~Corley and T.~Jacobson,
  %``Hawking Spectrum and High Frequency Dispersion,''
  Phys.\ Rev.\  D {\bf 54}, 1568 (1996)
  [arXiv:hep-th/9601073].
  %%CITATION = PHRVA,D54,1568;%%

\bibitem{Jacobson:2007jx}
  T.~Jacobson and R.~Parentani,
  %``Black hole entanglement entropy regularized in a freely falling frame,''
  Phys.\ Rev.\  D {\bf 76}, 024006 (2007)
  [arXiv:hep-th/0703233].
  %%CITATION = PHRVA,D76,024006;%%

\bibitem{thooft}
  G.~'t Hooft,
  %``On The Quantum Structure Of A Black Hole,''
  Nucl.\ Phys.\  B {\bf 256}, 727 (1985).
  %%CITATION = NUPHA,B256,727;%%

\bibitem{hzk}
  F.~He, Z.~Zhao and S.~W.~Kim,
  %``Statistical entropies of scalar and spinor fields in Vaidya-de Sitter
  %space-time computed by the thin-layer method,''
  Phys.\ Rev.\  D {\bf 64}, 044025 (2001);
%\bibitem{zl}
  Z.~A.~Zhou and W.~B.~Liu,
  %``Entropy calculation of a Kerr-Newman black hole via the thin film
  %brick-wall model,''
  Int.\ J.\ Mod.\ Phys.\  A {\bf 19}, 3005 (2004).
  %%CITATION = IMPAE,A19,3005;%%

\bibitem{li:plb02540}
  X.~Li,
  %``Black hole entropy without brick walls,''
  Phys.\ Lett.\  B {\bf 540}, 9 (2002)
  [arXiv:gr-qc/0204029];
  %%CITATION = PHLTA,B540,9;%%
%\bibitem{liu:membrane}
  %%CITATION = CPLEE,20,440;%%
  C.~Z.~Liu,
  %``Black hole entropies of the thin film model and the membrane model without
  %cutoffs,''
  Int.\ J.\ Theor.\ Phys.\  {\bf 44}, 567 (2005);
  %%CITATION = IJTPB,44,567;%%
%\bibitem{lhz}
  W.~B.~Liu, Y.~W.~Han and Z.~A.~Zhou,
  %``Black hole entropy inside and outside the brick wall,''
  Int.\ J.\ Mod.\ Phys.\  A {\bf 18}, 2681 (2003);
  %%CITATION = IMPAE,A18,2681;%%
%\bibitem{sl}
  X.~F.~Sun and W.~B.~Liu,
  %``Improved black hole entropy calculation without cutoff,''
  Mod.\ Phys.\ Lett.\  A {\bf 19}, 677 (2004).
  %%CITATION = MPLAE,A19,677;%%

\bibitem{kkp}   W.~Kim, Y.~W.~Kim and Y.~J.~Park,
  %``Entropy of the Randall-Sundrum brane world with the generalized
  %uncertainty principle,''
  Phys.\ Rev.\  D {\bf 74}, 104001 (2006)
  [arXiv:gr-qc/0605084];
  %%CITATION = PHRVA,D74,104001;%%
%\bibitem{kp}
  Y.~W.~Kim and Y.~J.~Park,
 %``Entropy of the Schwarzschild black hole to all orders in the Planck
  %length,''
  Phys.\ Lett.\  B {\bf 655}, 172 (2007)
  [arXiv:0707.2128 [gr-qc]];
  %%CITATION = PHLTA,B655,172;%%
%\bibitem{ko}
  W.~Kim and J.~J.~Oh,
  %``Determining the Minimal Length Scale of the Generalized Uncertainty
  %Principle from the Entropy-Area Relationship,''
  JHEP {\bf 0801}, 034 (2008)
  [arXiv:0709.0581 [hep-th]].
  %%CITATION = JHEPA,0801,034;%%

\bibitem{zrl}
  T.~Zhu, J.~R.~Ren and M.~F.~Li,
  %``Influence of Generalized and Extended Uncertainty Principle on
  %Thermodynamics of FRW universe,''
  Phys.\ Lett.\  B {\bf 674}, 204 (2009)
  [arXiv:0811.0212 [hep-th]];
  %%CITATION = PHLTA,B674,204;%%
%\bibitem{li:prd}
  Z.~H.~Li,
  %``Energy distribution of massless particles on black hole backgrounds with
  %generalized uncertainty principle,''
  Phys.\ Rev.\  D {\bf 80}, 084013 (2009).
  %%CITATION = PHRVA,D80,084013;%%

\bibitem{Ko:2010zz}
  Y.~Ko, S.~Lee and S.~Nam,
  %``Modified black hole thermodynamics with generalized uncertainty
  %principle,''
  Int.\ J.\ Theor.\ Phys.\  {\bf 49}, 1384 (2010).
  %%CITATION = IJTPB,49,1384;%%

\bibitem{Suyama:2009vy}
  T.~Suyama,
  %``Notes on Matter in Horava-Lifshitz Gravity,''
  arXiv:0909.4833 [hep-th].
  %%CITATION = ARXIV:0909.4833;%%

\bibitem{Liu:2001ra}
  W.~-B.~Liu, Z.~Zhao,
  %``The entropy calculated via brick-wall method comes from a thin
  %film near the event horizon,''
  Int.\ J.\ Mod.\ Phys.\  {\bf A16}, 3793-3803 (2001).

\bibitem{Garay:1994en}
  L.~J.~Garay,
  %``Quantum gravity and minimum length,''
  Int.\ J.\ Mod.\ Phys.\  A {\bf 10}, 145 (1995)
  [arXiv:gr-qc/9403008].
  %%CITATION = IMPAE,A10,145;%%

\bibitem{Chang:2001bm}
  L.~N.~Chang, D.~Minic, N.~Okamura and T.~Takeuchi,
  %``The Effect of the Minimal Length Uncertainty Relation on the Density of
  %States and the Cosmological Constant Problem,''
  Phys.\ Rev.\  D {\bf 65}, 125028 (2002)
  [arXiv:hep-th/0201017].
  %%CITATION = PHRVA,D65,125028;%%

\bibitem{Yoon:2007aj}
  M.~Yoon, J.~Ha and W.~Kim,
  %``Entropy of Reissner-Nordstrom Black Holes with Minimal Length Revisited,''
  Phys.\ Rev.\  D {\bf 76}, 047501 (2007)
  [arXiv:0706.0364 [gr-qc]].
  %%CITATION = PHRVA,D76,047501;%%

\end{thebibliography}

\end{document}